\documentstyle[12pt]{article}
\newcommand{\eqr}[1]{(\ref{#1})}
\newfont{\feff}{cmti10}
\def\ds{\displaystyle}
\def\undertext#1{\vtop{\hbox{#1}\kern 1pt \hrule}}
\def\ra{\rightarrow}

\def\d{\hbox{d}\,}
\def\bn{\beta N}
\def\bon{\beta_0 N}
\def\lr{\lambda_r(\bn)}
\def\lor{\lambda_r(\bon)}
\def\ep{\epsilon}

\def\e{{\,\rm e}\,}
\newcommand{\tr}[1]{\,{\rm tr}\,#1\,}

\def\pinta{\,{-\!\!\!\!\!\!\int_{-a}^{a}}}
\def\pint1{\,{-\!\!\!\!\!\!\int_{0}^{1}}}
\def\be{\begin{equation}}
\def\ee{\end{equation}}
\def\bea{\begin{eqnarray}}
\def\eea{\end{eqnarray}}
\def\eqref#1{\ref{(#1)}}
\begin{document}
\begin{flushright} TAUP-2012-92\\ December 1992 \\hep-th/9212090
\end{flushright}
\vskip2cm
\begin{center}
{\LARGE \bf Large-N quantum gauge theories}
\vskip.3cm
{\LARGE \bf in two dimensions}
\vskip1.5cm
{\large B.Rusakov}
\vskip.5cm
{\it School of Physics and Astronomy \\
Raymond and Beverly Sackler Faculty of Exact Sciences \\
Tel-Aviv University, Tel-Aviv 69978, Israel }
\vskip3cm
{\bf Abstract.}
\end{center}

\noindent
The partition function of a two-dimensional quantum gauge theory
in the large-$N$ limit is expressed as the functional integral
over some scalar field. The large-$N$ saddle point equation is
presented and solved. The free energy is calculated as the function
of the area and of the Euler characteristic. There is no non-trivial
saddle point at genus $g>0$. The existence of a non-trivial saddle
point is closely related to the weak coupling behavior of the theory.
Possible applications of the method to higher
dimensions are briefly discussed.

\newpage
It is well known that quantum gauge theories are exactly soluble in two
dimensions. It makes two dimensional model a useful instrument for
probing a new methods eventually intended to higher dimensions.
In this letter we develop a new large-$N$ approach for the gauge
theories, and apply it to the two-dimensional pure gauge theory.
We consider the model in the lattice formulation, given by K.Wilson
\cite{Wilson}. A convenient approach to this model at arbitrary $N$ is
the group-theoretical expansion proposed by A.A.Migdal \cite{Mig}.
This approach has been applied by the author to calculate loop
averages\footnote{All loop averages on a {\it plane} were
calculated by V.A.Kazakov and I.K.Kostov \cite{KK}. See also paper of
N.Bralic \cite{Br}.}
and the partition function of the model in two dimensions
at arbitrary finite $N$ \cite{Ru}. The results were expressed through
sums over irreducible representations of the gauge group.

The idea we develop below is that in the large-$N$ limit the
finite-$N$ signatures (parametrized by the $N$ numbers), can be replaced
by a scalar field, while the sum over all signatures
can be represented by the functional integral over this scalar field.

Using the exact results of \cite{Ru}, we realize this program in two
dimensions (we consider the case of arbitrary {\it orientable} surfaces).
We solve the large-$N$ saddle point equation by the method
described in ref.\cite{BIPZ} and calculate the free energy
as a function of the area and of the topological characteristics
of the manifold.

In conclusion we discuss the difference between our approach and that
of ref.\cite{GW} and the possibility of applying both of them
together to  higher-dimensional models.

Thus, we consider the lattice model \cite{Wilson}
\be S=\bon\sum_{f}\tr[U_f+U_f^\dagger] ,  \label{Mod} \ee
where sum goes over all faces (fundamental polygons) $f$ of the
two-dimensional lattice and $\beta_0$ is the lattice coupling constant.

Following \cite{Mig} we expand the contribution of each face
over irreducible representations, $r$, of the (compact) gauge group :
\be \e^{\bon\tr[U_f+U_f^\dagger]}=\sum_{r}\d_r\lor\chi_r(U_f) ,
\label{ex} \ee
where $\chi_r(U)$ and $\d_r=\chi_r(I)$ are characters and dimensions
of $r$'s respectively. The coefficients of the expansion \eqr{ex}
\be \lor=\int DU\e^{\bon\tr[U+U^\dagger]}\chi_r(U)
\label{lr} \ee
have the following asymptotic behavior in the continuum limit $\ep\ra 0$
($\ep $ is the area of the face and $\beta =\ep\beta_0 $) :
\be  \lr ~\sim ~ \exp\left(- {{C_2(r)}\over{2\bn}}\ep \right)  ,
\label{lras}  \ee
where $C_2(r)$ is the eigenvalue of the quadratic Casimir operator.

In ref.\cite{Ru} the following expression for the
partition function has been derived\footnote
{In ref.\cite{Ru} this formula was obtained actually for
$U(N)$ and $SU(N)$ gauge groups but, as it can be easily proven, it
holds for any compact gauge group.}:
\be
Z=\sum_{r}\d_r^\eta \exp\left(-{{C_2(r)A}\over{2\bn}}\right)
\label{Z} \ee
where $A$ is the area of the surface and $\eta=2-2g$ is the
Euler characteristic, with $g$ being a number of the handles (genus).
It has been argued in \cite{Ru} that $Z=1$ in the case of the surface
with holes.

Now we substitute in eq.\eqr{Z} an explicit expressions for dimensions,
\be
\d_r=\prod_{i<j}^{N}\left(1+{{n_i-n_j}\over{j-i}}\right) ,\label{dim}\ee
and Casimir eigenvalues,
\be C_2(r)=\sum_{k=1}^{N}\left(n_k^2 + n_k(N-2k+1)\right) ,
\label{Cas} \ee
where $n_k$'s are parameters of signature:
$r=\left\{n_1,\dots,n_N\right\}$ obeying the dominance condition
$n_1\geq\dots\geq n_N$. Hence, eq.\eqr{Z} reads
\be Z=\sum_{n_1}^{\infty}\dots
\sum_{n_N}^{\infty}\prod_{k=1}^{N-1}\theta(n_k-n_{k+1}) \e^S , \ee
\be S=-{A\over{2\bn}} \sum_{k=1}^{N}\left(n_k^2 + n_k(N-2k+1)\right)
+{{\eta}\over 2}\sum_{i\neq j}
\log\left(1+{{n_i-n_j}\over{j-i}}\right) , \ee
where the step function, $\theta(n)=1$ if $n\geq 0$ and
$\theta(n)=0$ if $n<0$, realizes the dominance condition for
signatures.

For large $N$, we introduce a continuum time,
$0\leq x={k\over N}\leq1$, and replace sum over $n_k$'s
by the path integral over the scalar field $n(x)$ :
\be Z=\int\prod_{1\leq x\leq 0} dn(x) ~ \e^S
\label{Zn} \ee
\be S={{N^2}\over 2} \int_0^1 \!\!\! dx
\left\{-{A\over{\beta}}\left(n^2(x)+n(x)(1-2x)\right)+
\eta \pint1 \!\! dy \log\left(1+{{n(x)-n(y)}\over{y-x}}\right)
\right\} . \label{act} \ee
We omit here the step function since its contribution to the
action is of order $N$ (i.e., $1/N$ with respect to expression
\eqr{act}) \footnote{More detailed discussion of this point will
be given in \cite{Ru2}.}.

Now, we calculate \eqr{Zn} using the saddle point method.
First, we replace $n(x)$ by the new field
$\phi(x)=n(x)-x+{1\over 2}$. Then, the saddle
point equation is
\be 2\xi\phi(x)=\pint1 {{dy}\over{\phi(x)-\phi(y)}}
\:\;\; ; \;\;\;\;\xi={{\ds A}\over{\ds \beta\eta}} \;\; .
\label{sadeq}\ee
Introducing the density \be \rho(\phi)={{dx}\over{d\phi}}\label{den}\ee
which should be positive, even and normalized to
\be \int_{-a}^{a}\!\! d\lambda\;\rho(\lambda)=1 , \label{norm} \ee
we rewrite eq.\eqr{sadeq} as equation for $\rho$:
\be 2\xi \lambda=\pinta{{d\mu \rho(\mu)}\over{\lambda-\mu}}
\:\;\; ; \;\;\;\; |\lambda|\leq a \; . \label{Eq}\ee
The solution of eq.\eqr{Eq} is
\footnote{The method of solution of such an equations can be found
in ref.\cite{BIPZ}.}
\be \rho({\lambda})={2\over{\pi}} \sqrt{\xi(1-\xi\lambda^2)}
\label{rho}\ee
with
\be a={1\over{\sqrt{\xi}}} . \label{bound} \ee
Now, we transform \eqr{act} into
\be S={{N^2}\over 2} \left\{ {A\over{12\beta}}+{3\over 2}\eta
-{A\over{\beta}}\int_{-a}^{a}\!\! d\lambda\;\rho(\lambda) \left(
\lambda^2 -\xi^{-1}\pinta \!\!d\mu\;\rho(\mu)\log|\lambda-\mu|
\right) \right\} . \label{act2}\ee
Then, integrating \eqr{Eq} with respect to $\lambda$ and
defining the free energy as $F={2\over{N^2}}\log Z$ we have
\be F= {A\over{12\beta}}+{3\over 2}\eta
+\eta\int_{-a}^{a} \!\!d\lambda\;\rho(\lambda)\log|\lambda|
\ee
and, finally,
\be F= {A\over{12\beta}}
+\eta \left(1+{1\over 2}\log{{\eta\beta}\over{4A}} ~\right) \; .
\ee

We see from \eqr{rho},\eqr{bound} that there is no non-trivial large-$N$
saddle point for genus $g > 0$ ($\xi$ and $\eta$ become negative).
The appearance of this phenomenon is clear already from formula \eqr{Z}.
A saddle point exists only when the topological (entropy) term in the
effective action (this term arises from powers of $\d_r$'s) acts in
the opposite direction to the area (energy) term.

In the case of a torus ($g=1$) there is no topological term.
The corresponding free energy is $F= {\ds A\over{\ds 12\beta}}$.

At higher genera, $\eta < 0$,
there are only negative powers of $\d_r$ in \eqr{Z} and the
topological term acts in the same direction as the area term.
The partition function \eqr{Z} in the large-$N$ limit is then
dominated by the trivial representation ($\d_r=1$, ~$C_2(r)=0$) and
corresponding free energy equal to zero.

The non-trivial large-$N$ behavior occurs only for a sphere ($g=0$).
In this case, the positive powers of $\d_r$'s give a positive
contribution to the effective action and compensate (at the saddle
point) the negative contribution of the area term.

To summarize, we write the free energy of the large-$N$ quantum gauge
theory on an orientable surface with $g$ handles and with $h$ holes,
\be F=\left\{ \begin{array}{ll}
{\ds A\over{\ds 12\beta}} + 2 + \log{{\ds \beta}\over{\ds 2A}}
&,~~~{\rm sphere} :  g=0 ~{\rm and} ~ h=0   \\  \\
{\ds A\over{\ds 12\beta}}&,~~~{\rm torus} :  g=1 ~{\rm and} ~ h=0 \\ \\
0 &,~~~ g>1 ~{\rm or~(and)} ~ h > 0
\end{array}\right.\ee

This formula sums all planar diagrams of the quantum gauge
theory in two dimensions.

Note, that the non-trivial large-$N$ saddle point exists only
when in the weak coupling limit ($\beta\ra\infty$ in our notation)
the free energy is divergent. Apparently, this is the case in
higher dimensions and we can hope that our method will be
relevant there.

Possible application of our approach to higher dimensions is intimately
connected to the problem of the (one-link) integration over unitary
matrix which is already not so simple as in two dimensions, even at
large $N$. This problem has been studied by D.Gross and E.Witten in
\cite{GW} and a third order phase transition was found. The technical
reason for such a phase transition is the unitarity constraint:
the eigenvalue density (the analogue of our quantity \eqr{den}),
\be \rho({\alpha})={2\over{\pi}}\beta \cos(\alpha)
\sqrt{{1\over{2\beta}} - \sin^2(\alpha)}  \ee
(where $\alpha(x)$ is the scalar field coming from the unitary
matrix eigenvalues) depends on the functions bounded by 1,
and, consequently, there are two different types
of behavior for coupling constant $\beta > {1\over 2}$ and for
$\beta < {1\over 2}$.
{}From the point of view of our approach this phenomenon seems a lattice
artifact, since considering the model, where integration over
unitary matrices (over $\alpha$'s at large $N$) is performed from very
beginning \cite{Ru}, we do not realize a phase transition with respect
to $\beta$.

To conclude, the following proposal for higher dimensions can be made.
The large-$N$ limit could be described in terms of both a scalar field
$n(x)$ arising from the signatures and a scalar field $\alpha(x)$
arising from eigenvalues of the unitary matrix, with a properly defined
integration over $\alpha$'s.
Introducing two scalar fields may seems like an unnecessary
complication, but
this complication could give more freedom to solve the problem.
\bigskip

\noindent {\bf Acknowledgements.}

\noindent
I am grateful to A.A.Migdal for drawing my attention to this problem and
for discussion. I thank also M.Karliner, N.Marcus, J.Sonnenschein and
S.Yankielowicz for discussion and D.V.Boulatov for valuable comments.
This research has been supported in part by the Basic Research Foundation
administered by the Israel Academy of Sciences and Humanities, by a
grant from the United States--Israel Binational Science Foundation (BSF),
Jerusalem, Israel, and also by the Israel Ministry of Absorption.


\begin{thebibliography}{99}
\small  \addtolength{\itemsep}{-6pt}

\bibitem{Wilson}
K.Wilson, {\sl Phys.Rev.} {\bf D10}, 2445 (1974).
%
%
\bibitem{Mig}
A.A.Migdal, {\sl ZhETF} {\bf 69} (1975) 810
({\sl Sov.Phys.JETP} {\bf 42} 413).
%
\bibitem{KK}
V.A.Kazakov, I.K.Kostov,
{\sl Nucl.Phys.} {\bf B176} (1980) 199;\\
V.A.Kazakov, {\sl Nucl.Phys.} {\bf B179} (1981) 283.
%
\bibitem{Br}
N.Bralic, {\sl Phys.Rev.} {\bf D22} (1980) 3090.
%
\bibitem{Ru}
B.Rusakov, {\sl Mod.Phys.Lett.} {\bf A5} (1990) 693.
%
\bibitem{BIPZ}
E.Brezin, C.Itzykson, G.Parisi, J.B.Zuber,
{\sl Comm.Math.Phys.} {\bf 59} (1978) 35.
%
\bibitem{GW}
D.J.Gross, E.Witten,
{\sl Phys.Rev.} {\bf D21} (1980) 446.
%
\bibitem{Ru2}
B.Rusakov, {\sl in progress.}
%
\end{thebibliography}
\end{document}